\documentclass[journal=jacsat,manuscript=article]{achemso}

\usepackage[version=3]{mhchem} 
\usepackage{siunitx}
\usepackage{textcomp}

\author{Evgeniy M. Myshakin}
\email{Evgeniy.Myshakin@netl.doe.gov}
\affiliation[]{National Energy Technology Laboratory, 626 Cochrans Mill Road, Pittsburgh, Pennsylvania 15236, United States}
\alsoaffiliation[]{URS Corporation, P.O. Box 618, South Park, Pennsylvania 15129, United States}
\author{Meysam Makaremi}
\affiliation[]{National Energy Technology Laboratory, 626 Cochrans Mill Road, Pittsburgh, Pennsylvania 15236, United States}
\alsoaffiliation[]{University of Pittsburgh, Department of Chemistry, Pittsburgh, Pennsylvania 15260, United States}
\author{Vyacheslav N. Romanov}
\affiliation[]{National Energy Technology Laboratory, 626 Cochrans Mill Road, Pittsburgh, Pennsylvania 15236, United States}
\author{Kenneth D. Jordan}
\affiliation[]{National Energy Technology Laboratory, 626 Cochrans Mill Road, Pittsburgh, Pennsylvania 15236, United States}
\alsoaffiliation[]{University of Pittsburgh, Department of Chemistry, Pittsburgh, Pennsylvania 15260, United States}
\author{George D. Guthrie}
\affiliation[]{National Energy Technology Laboratory, 626 Cochrans Mill Road, Pittsburgh, Pennsylvania 15236, United States}

\title[An \textsf{achemso} demo]
  {Molecular Dynamics Simulations of Turbostratic Dry and Hydrated
Montmorillonite with Intercalated Carbon Dioxide}

\abbreviations{IR,NMR,UV}

\begin{document}
\begin{abstract}
Molecular dynamics simulations using classical force fields were carried out to study energetic and structural properties of rotationally disordered clay mineral-water-CO$_2$ systems at pressure and temperature relevant to geological carbon storage. The simulations show that turbostratic stacking of hydrated Na- and Ca-montmorillonite
and hydrated montmorillonite with intercalated carbon dioxide is an energetically demanding process accompanied by an increase in the interlayer spacing. On the other hand, rotational disordering of dry or nearly dry smectite systems can be energetically favorable. The distributions of interlayer species are calculated as a function of the rotational angle between adjacent clay layers.
\end{abstract}

\section{Introduction}
Carbon capture and storage technologies offer an important option for reducing $CO_2$ emissions and mitigating global climate change\cite{NETL}. The technology of geologic $CO_2$ storage involves injection of supercritical $CO_2$ ($scCO_2$) into deep geologic formations overlain by sealing rocks that prevent buoyant $CO_2$ from migrating upward and out of the storage reservoir. The ability of cap rocks to retain injected $CO_2$ depends on their ability to maintain low permeability\cite{Allen2005}. Caprocks are often composed of shale or mudstone enriched with
swelling clay minerals that may expand or contract upon interaction with $scCO_2$ that, in turn, could impact seal permeability, $CO_2$ mobility, or both\cite{Abdou2010,Ilton2012,Schaef2012}. 
Intercalation of $CO_2$ into the interlayer of swelling clay can cause geo-mechanical
stress and affect the integrity of cap rocks and the ability of geological formations to contain stored $CO_2$\cite{Schaef2012,Kwak2011,Rother2012}. The integrity of the cap rock is important because $CO_2$, being more buoyant than saline water and oil, will tend to migrate above these relatively immiscible fluids. Moreover, swelling clay minerals may also provide sites for sorption of $CO_2$\cite{Schaef2012,Romanov2013,Romanov2010,Loring2012,Giesting2012} or environments for its transformation into carbonates\cite{Hur2013}.\newline
Swelling clay minerals generally fall into the smectite group,
which consists of a group of layered aluminosilicate mineral
species with a wide range in compositional variability. We have
focused our study specifically on montmorillonite (MMT), for
which the central dioctahedral sheet is composed of
octahedrally coordinated aluminum (Al) atoms and the
adjacent sheets contain tetrahedrally coordinated silicon (Si)
atoms. These sheets comprise a 2:1 or tetrahedral\_octahedral\_tetrahedral (TOT) layer. In MMT, the TOT layers are
negatively charged due to substitution of divalent metals (e.g.,
$Mg^{2+}$ for $Al^{3+}$ in the octahedral sites). The negative charge of
the TOT layers is counterbalanced by interlayer cations (e.g., $Na^+$, $K^+$, $Ca^{2+}$, etc.) that can exhibit a variety of hydration states, causing expansion or contraction of the interlayer distance of
the clay, depending on the relative humidity\cite{Fu1990} Recent X-ray diffraction (XRD), excess sorption, and neutron diffraction studies showed that the spacing between the mineral layers (basal d-spacing) of Na-rich-MMT expands upon interaction with gaseous $CO_2$ and $scCO_2$ and that the degree of expansion depends on the initial water content in the interlayers\cite{Rother2012,Giesting2012,Giesting2012a}. Those measurements indicated that the largest expansion is accompanied by an increase in the basal d-spacing from 11.3 to 12.3 \AA, the number corresponding to a stable monolayer (the 1W hydration state characterized with basal d$_{spacing}$s in the range of 11.5-12.5 \AA; upon incorporation of more water, the d-spacing increases to 14.5-15.5 \AA\space to the
stable 2W hydration state where water forms a bilayer structure)\cite{Fu1990,Marry2002}. Interaction of anhydrous $scCO_2$ with smectite clay in the 2W and higher hydration states may lead in a collapse of the d-spacing to that of the 1W state\cite{Ilton2012,Schaef2012}. However, interaction of Na-exchanged Wyoming montmorillonite (SWy) and Ca-exchanged Texas montmorillonite (STx) samples with variably wet $scCO_2$ (2-100 \% saturation of $H_2O$) induces swelling even to the values equal to the 3W state with a d-spacing of 18.8 \AA \cite{Ilton2012,Schaef2012}. 
Thus, dry $scCO_2$ injected in a target
reservoir has the capacity to dehydrate clay and to promote
fracturing of cap rocks. On the other hand, after $CO_2$ becomes
saturated by brine, it can induce further swelling of clay. Loring
et al.\cite{Loring2012} confirmed intercalation of $CO_2$ by means of NMR
spectroscopy and attenuated total reflection infrared spectroscopy.
Using diffuse-reflectance infrared spectroscopy, Romanov\cite{Romanov2013}
reported a red shift of the characteristic fundamental
frequency of $CO_2$ trapped in SWy and STx samples. The source
of that shift was attributed to interaction of the intercalated
$CO_2$ molecules with dipoles of water molecules.\cite{Myshakin2013}
\\Botan et al.\cite{Botan2010} carried out Monte Carlo (MC) and molecular
dynamics (MD) simulations of $CO_2$ intercalation into Na-MMT using a force field for clay from Smith\cite{Smith1998} and the SPC\cite{Berendsen}
and EPM2\cite{Harris1995} models for water and carbon dioxide, respectively.
In line with experimental data,\cite{Ilton2012,Schaef2012,Rother2012,Giesting2012,Giesting2012a} their simulations
showed that hydrated clay intercalates $CO_2$ and that the
thermodynamically stable structures are characterized with
basal d-spacings corresponding to the 1W and 2W hydration
states. Other modeling studies using different force fields also
point out that $CO_2$ molecules can exist in the interlayers of clay
minerals.\cite{Yang2005,Peng2007,Cole2010,Krishnan2013} Krishnan et al.\cite{Krishnan2013} recently reported information
on the molecular-scale structure and dynamics of interlayer
species in MMT-$CO_2$ systems. An excellent
review of the recent advances in molecular modeling of $CO_2$-brine-mineral interactions is given in ref \cite{Hamm2013}.\newline
Another major feature of swelling clay minerals is related to rotational disorder inherently present in natural samples and manifesting itself through turbostratically stacked clay layers.\cite{Guthrie1998,Viani2002,Lutterotti2010,Moore1989} 
Turbostratic disorder, a disorder in which different layers have different rotations with respect to an axis, is commonly found in naturally occurring samples of montmorillonite.\cite{Lutterotti2010,Moore1989} 
To reconcile apparent discrepancies between high resolution transmission electron microscopy (HRTEM) images and powder X-ray diffraction (XRD),\cite{Veblen1990,Guven1973} Guthrie and Reynolds\cite{Myshakin2013} offered a model in which adjacent TOT layers of smectite are turbostratically stacked within $\sim 2-10$ degree rotation of each other.
\cite{Veblen1990,Guven1973,Mering1967, Reynolds1992} In this model, rotation of adjacent TOT layers results in breaking of coherency between the ditrigonal rings in the silica sheets on either side of the interlayer, thereby changing the framework structure that bounds the interlayer region. Instead of having a coherent alignment of ditrigonal rings on either side of the interlayer region, rotational disorder results in a Moire pattern with a periodic variation in alignment
of ditrigonal rings across the interlayer. In the simulations,
smectite models are assumed to be perfectly oriented, and any
impacts of rotational disorder are generally ignored. However,
understanding interactions between interlayer species and
rotationally disordered clay systems is mandatory to properly
predict the behavior of geological formations and cap rocks
under carbon dioxide invasion.
In the present study, classical molecular dynamics
simulations are used to investigate the process of rotational
disordering in MMT at various water and carbon dioxide
contents. The simulations were carried out at P-T conditions
relevant to geological formations for $CO_2$ storage. The main
focus of the study is placed on comparison of energetic and
structural changes associated with deviation of the clay layers
from their ideally stacked positions. In addition, one-dimensional
density profiles and two-dimensional density maps are
engaged to study the distributions of interlayer species as a
function of the rotational angle between adjacent TOT layers.

\section{Computional Details}
\subsection{Classical Force Field Simulations}
The force field calculations were carried out using the GROMACS package.\cite{Gromacs} For the clay system, the Clayff force field\cite{Cygan2004} that consists of
nonbonded (electrostatic and van der Waals) terms parametrized
for use with layered minerals was employed. For atoms i and j separated by a distance $r_{ij}$, the pairwise energy is given by
\begin{equation}
\ E_{ij} = \frac{q_iq_je^2}{4\pi\epsilon_0r_{ij}} + 4\epsilon_{ij}[(\frac{\sigma_{ij}}{r_{ij}})^{12} - (\frac{\sigma_{ij}}{r_{ij}})^{6}]
\end{equation}
where $q_i$ is the charge on atom i, $\epsilon_0$ is the vacuum permittivity,
and $\epsilon_{ij}$ and $\sigma_{ij}$ are the Lennard-Jones (LJ) energy and distance parameters, respectively. The flexible SPC model\cite{Berendsen} was used for
the water molecules placed in the interlayer space and for the
layer hydroxyl groups. For $CO_2$, a recently developed flexible
potential including intramolecular bond stretch and angle bend
was used.\cite{Cygan2012} The general expression for the total potential
energy is
\begin{equation}
\ E_{total} = E_{Coul} + E_{Vdw} + E_{stretch} + E_{bend}
\end{equation}
where harmonic potentials are used for the bond stretch and
angle bend terms.
\\The Lorentz ˆ'Berthelot mixing rule\cite{Allen1989} was used to obtain the
LJ parameters for interactions between unlike atoms. It is worth
mentioning that in an earlier paper\cite{Myshakin2013} we showed that the
description of $CO_2$ in liquid water using the force fields
described above closely reproduces that obtained with
simulations using "polarization-corrected" LJ parameters for
the unlike pair interactions in water/carbon dioxide mixtures.\cite{Vlcek2011}
The authors of ref \cite{Vlcek2011} used the SPC/E and EPM2 force fields to
simulate $CO_2$ in liquid water and to accurately reproduce the
experimental solubilities.
\\In the present study, the simulations were performed under
periodic boundary conditions (PBC) and used the particle-
particle particle-mesh (PPPM) Ewald method to treat long range
electrostatics.\cite{Allen1989} The cutoff radii for the nonbonded van
der Waals interactions and for the Ewald summation of the
electrostatics were chosen to be 11 \AA, with switching distances
starting from 10 \AA. Because of the use of cutoffs for the LJ
interactions, long-range dispersion corrections for energy and
pressure were applied.\cite{Gromacs} The leapfrog algorithm\cite{VanGunsteren1988} was used to
update positions every 0.5 fs.
\\Simulations of the turbostratically disordered clay layers is
not a trivial task. Rotation of one clay layer relative to another
destroys the periodicity of the systems which is problematic in
simulations using PBC. To preserve periodicity, the clay
systems represent a rectangular box with uneven sizes of
alternating clay layers (designated as "small" and "large"). Thus,
within a chosen angular range, rotation of small layers proceeds
within the boundaries of the simulation box defined by large
ones. This approach allows us to use PBC and to provide
external clay surfaces (clay edges) and the interstitial space
between the alternating layers accessible to the interlayer
species. The interstitial space is produced because of gaps
between adjacent replicas of small layers. In turbostratic clay
systems, the clay layers are stacked in the z direction, so the
rotation of a small layer relative to a large one occurs in the xy
plane around the axis connecting geometrical centers of the
layers and perpendicular to the internal clay surfaces. The [100]
and [010] edges introduced by the presence of the small layers
are determined by the structure of the unit cell. Because the
Clayff force field\cite{Cygan2004} contains no bonding terms except for the
hydroxyl groups, there are no broken chemical bonds at the
edges.
\\The general chemical formula used for sodium montmorillonite
is $Na_xMg_xAl_{3-x}Si_4O_{10}$-$(OH)_2·nH_2O$, where the layer
charge resides on the octahedral sheet (tetrahedral substitutions
such as $Al^{3+}$ for $Si^{4+}$ can also occur in natural samples but they
are not considered here). In addition to sodium ions, calcium
and potassium (dry clay systems only) ions were also used as
interlayer ions in the simulation models. The MMT structural
model was created by replicating a pyrophyllite unit cell with an
isomorphic octahedral $Al^{3+}/Mg^{2+}$ substitution to produce a
rectangular $22 \times 14 \times 1 (18 \times 10 \times 1)$ supercell, in which the first set of numbers designates the size of the large layer and the
second one (in parentheses) designates the size of the smaller
layer. The initial dimensions of the simulation box in the x and
y directions were 114.4 and 128.8 \AA, respectively. The
dimension in the z direction varies depending on the interlayer
composition. The negative charge introduced by the substitutions
is compensated by 366 sodium ions (or 183 calcium
ions) residing in the interlayers.
\\The stoichiometry is $Na_{0.75}Mg_{0.75}Al_{3.25}(OH)_4(Si_4O_{10})_2$, with
a layer charge of 0.75 per $O_{20}(OH)_4$. This results in a total of
19520 atoms constituting the clay phase. Additional simulations
were performed for the dehydrated systems using a simulation
box doubled in size in the z direction to ensure that there is no
size effect on the results of simulations. In these tests, the
systems are described with a $22 \times 14 \times 2 (18 \times 10 \times 2)$
supercell. In addition, a $22 \times 12 \times 2 (18 \times 10 \times 2)$ supercell was used to confirm independence of reported below results on
the size of the interstitial space. MD simulations using a $8 \times 4 \times
4$ supercell were also carried out for dehydrated MMT systems
to estimate equilibrium $d_{001}$-spacing values for clay systems
providing exposure of the ions to internal surfaces only (i.e.,
without edges).
\\The water/$CO_2$ composition in the interlayer region is
designated as X-Y, where X and Y are, respectively, the
numbers of water and $CO_2$ molecules per unit cell. There are
four types of interlayer compositions used in the simulations:
the first one (0-0) contains only interlayer ions without water
and represents the dehydrated clay phase; the second (X-0)
contains X water molecules per unit cell; the third (0-Y)
includes Y carbon dioxide molecules per unit cell only, and the
fourth (X-Y) contains both X water and Y $CO_2$ molecules per
unit cell. The maximum number of atoms engaged in the
simulations was 40214 in the case of the clay system (Na-
MMT with a $22 \times 14 \times 2$ $(18 \times 10 \times 2)$ supercell) with the
14-0 composition.
\\The initial positions of the ions, water, and carbon dioxide
molecules were chosen randomly in a plane at the middle of the
interlayers. They are placed nearly equidistantly from each
other, and the plane dimensions are equal to the x and y
dimensions of the large layer. For the dry, $CO_2$ only, and
hydrated clay systems, multiple independent simulations
starting from different initial structures were performed to
ensure that the computed trends were consistent and were not
affected by different initial positions. Because of the gaps
between small layers, direct comparison of the energetic and
structural parameters with those computed for systems without
explicit presence of edges is not straightforward. Particularly,
the numbers of water molecules per unit cell corresponding to
the $d_{001}$-spacings at the stable hydration states (1W, 2W,
etc.)\cite{Krishnan2013,Cygan2012} are not the same for the systems considered in this work because water molecules reside in both the interlayer and
the interstitial space. For the X-Y compositions, the
concentrations of $CO_2$ and $H_2O$ in the interlayer and the
interstitial space could vary and depend on the initial positions
of the species. However, regardless of that, the relative energetic
and $d_{001}$-spacing trends as a function of the rotational angle ($\theta$)
remain the same. For the X-Y compositions, the results are
reported using the initial distribution of the species similar to
that described above.
\\MMT with intercalated species was initially relaxed using
MD simulations in the NVT ensemble carried out for 50 ps at T
= 348.15 K. Subsequent equilibration was conducted for 1 ns
using the NPT ensemble with the weak coupling Berendsen
thermostat and barostat\cite{Berendsen1984} at T = 348.15 K and P = 130 bar,
conditions close to those existing in subsurface geological
reservoirs and cap rocks and also close to those used in $CO_2$
intercalation experiments in clays carried out at National
Energy Technology Laboratory (NETL).\cite{Romanov2013} That step was
followed by 20 ns production runs in the NPT ensemble at
the same P-T conditions with semi-isotropic pressure coupling
permitting the z-dimension to fluctuate independently from the
x and y directions. Pressure was controlled by a Parrinello-Rahman barostat\cite{Parrinello1981,Nose1983} with a relaxation time of 4 ps, and the
temperature was controlled by a Nose-Hoover thermostat\cite{Nose1984,Hoover1985}
with a relaxation time of 2 ps. The analysis of different terms
contributing to the net potential energy was performed using
the equilibrated clay structures. For that purpose, MD
simulations were performed over 300 ps in the NVT ensemble
using a Nose-Hoover thermostat\cite{Nose1984,Hoover1985} with a relaxation time of
2 ps.\newline
\begin{figure}[h!]
\centering
\includegraphics[scale=0.5]{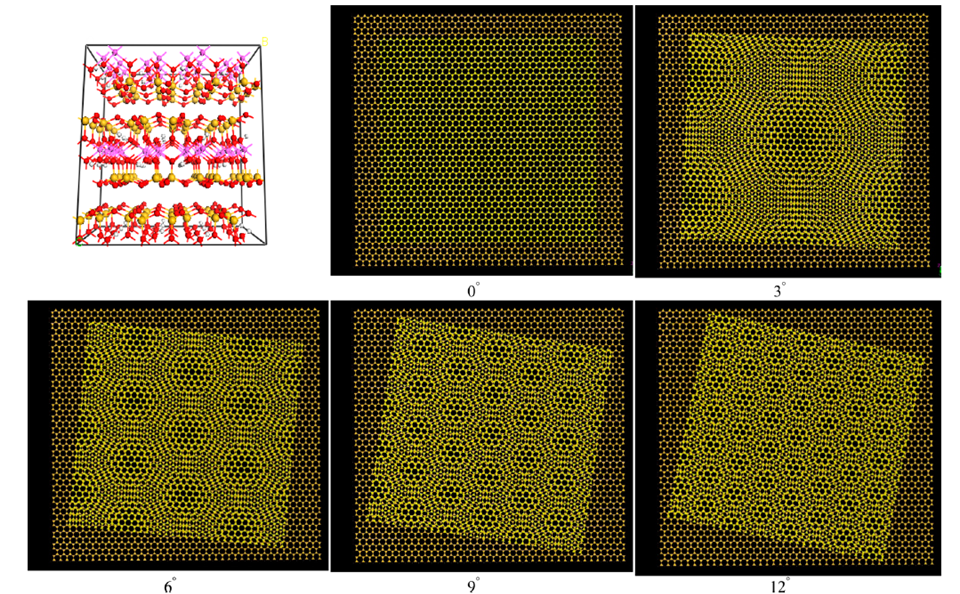}
\caption{The [100] edge face of the 2:1 smectite clay structure (pyrophyllite) and Moire patterns formed by two adjacent basal surfaces at 0-12\textdegree\space
(from the [001] view). Color designation: red balls, oxygen; yellow, silicon; white, hydrogen; cyan, aluminum.}
\end{figure}

\subsection{Rotational Pattern Model}
Figure 1 shows a schematic diagram of two adjacent tetrahedral sheets formed
by basal oxygens and silicon atoms starting from perfectly
juxtaposed sheets and with rotation angles of 3, 6, 9, and 12\textdegree\space
together with the 3D smectite structure (pyrophyllite was taken
for simplicity with the expanded size of the sheets to fully
capture the pattern at $\theta$ = 3\textdegree). For perfectly aligned sheets, two
ditrigonal rings of silicon tetrahedra linked by shared basal
oxygens locate above each other and create a cavity shown in
Figure 2a. Rotational mismatch results in complex cavities
reflecting a variation in alignment of the ditrigonal rings across
adjacent clay sheets. The variation in cavity structure is
periodic, forming a Moire pattern with concentric circles of
partially aligned ditrigonal rings separated by regions of no
alignment (Figures 1 and 2b). For small rotational shifts, the
distances between the centers of the circles along a layer are given by
\begin{equation}
\ d_{\mathit{MP}} = \frac{d_a}{2 sin(\frac{\theta}{2})}
\end{equation}
where $d_{\mathit{MP}}$ is the distance between centers of two circles in the
Moire pattern, $d_a$ is the distance between centers of ditrigonal
rings of the same layer and equal to the a lattice, and $\theta$ is the
angle by which one layer is rotated relative to another.
Assuming a $d_a$ value equal to 5.20 /AA (a typical distance for the a
lattice constant of MMT\cite{Tsipursky1984}) the calculated $d_{\mathit{MP}}$ values are 99, 50,
33, and 25 \AA\space for $\theta$ = 3\textdegree, 6\textdegree, 9\textdegree, and 12\textdegree, respectively. Those numbers dictate that to capture the patterns, a sufficiently large
model is required. To fulfill this requirement, the size of the
simulation box was chosen as described in section 2.1. The area
of one unit cell (Figure 2b; inside the blue arrows) formed by
the rotational pattern is further determined as
\begin{equation}
\ S_{\mathit{MP}} = d_{\mathit{MP}}^2 sin(\pi/3) = \frac{\sqrt{3}}{2}\frac{d_a^2}{(2 sin(\frac{\theta}{2}))^2 }
\end{equation}
The area of the concentric circle itself is computed as
\begin{equation}
\ S_{IS} = \pi d_{IS}^2 = \frac{\pi d_a^2}{(2F sin(\frac{\theta}{2}))^2}
\end{equation}
where $d_{IS}$ is the radius within which the shifted ditrigonal rings
are still viewed as forming concentric circles, $\theta$ has the same
meaning as in eq 3, and F is a factor determining the fraction of
the unit cell that belongs to the circles. This formula is valid
until $d_{IS}$ is larger than the lattice parameter a. Assuming that the
concentric circles occupy 1/3 of the unit cell area, the limiting
value of the $\theta$ angle is 17.4\textdegree, after which the rotational pattern
disappears. Thus, the chosen range of the rotational angles, [0-12\textdegree], passes through the various Moire patterns and reflects the
interval of angles consistent with TEM observations\cite{Moore1989,Veblen1990,Guven1973,Mering1967}
Another implication of this model is that the ratio of the unit
cell area to the concentric circle area does not depend on $\theta$.
This means that the fraction of the concentric circles remains
fixed during the limited range of rotation considered. From the
mathematical model presented above, it follows that the
number of cavities formed by basal oxygens of the adjacent
layers (Figure 2a) is constant within the rotation range,
although their distribution is varied. By definition, that number
is smaller than the number of (undistorted) cavities for the 0\textdegree\space
case, in which clay layers are perfectly aligned.
\newline
\begin{figure}[h!]
\centering
\includegraphics[scale=0.5]{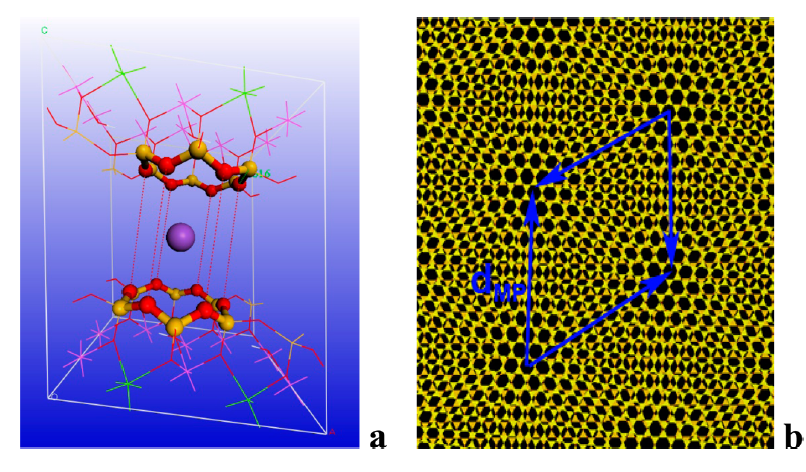}
\caption{(a) A "cavity" formed by basal atoms of adjacent clay layers and (b) the unit cell of the Moire pattern.}
\end{figure}
\subsection{Methods To Study Rotational Disordering}
The question of the time scales of rotational motion of clay layers is
most intriguing. On one hand, the clay minerals in a geological
formation could be exposed to geo-mechanical stress for
decades, which essentially implies nearly equilibrium conditions
for interlayer species during rotation. On the other hand, local
perturbations of a subsurface geological reservoir due to $CO_2$
injection could create mechanical forces acting on clay-rich
deposits to impose rotational motion of the layers at
non-equilibrium conditions. Hence, our simulation approaches
stem from the intention to simulate both rotational disordering
occurring on geological time scales and during injection of a
mobile phase into porous media of target formations. The
method mimicking geological conditions deals with rotated clay
systems with predetermined $\theta$ angles, at which interlayer
species are equilibrated. The second method rotates alternating
clay layers around a fixed axis with a constant angular velocity.
Details of the two methods are given below.
\subsubsection{Position Restraining} 
In this approach, the rotationally
disordered and 0\textdegree\space systems were initially prepared as described in section 2.1. The clay systems were equilibrated at $\theta$ = 0\textdegree, 3\textdegree, 6\textdegree, 9\textdegree, and 12\textdegree. The production runs with various compositions
of the interlayer species were carried out using the procedure
described above. To keep the atomic positions of the clay phase
at the predetermined angles, the atoms were harmonically
restrained in the XY plane at their reference positions using a
force constant of 5000 kJ/mol.nm2. This approach provides no
translational motion of the clay layers relative to each other.
Those degrees of freedom might be important for equilibration.\cite{Kutzner2011} The Z coordinates remained unconstrained, allowing the
simulation box to adjust its dimension in that direction during
the simulations. This approach provides equilibration of
interlayer species in the turbostratic clay systems (monitored
using constancy of potential energies and $d_{001}$-spacings over
simulation time) at the rotational angles of interest.
\subsubsection{Enforced Rotation}
Various methods for enforcing the rotation of subsets of atoms have been reported by Kutzner et al.\cite{Voora2011} In the approach engaged in this work, a force is imposed
on the group of atoms constituting the small clay layer by
means of rotating a reference set of atomic positions, $y_i^0$
(coinciding with initial atomic positions of the small clay
layer), at a constant angular velocity, $\omega$, around a fixed axis
defined through a geometrical center of the clay layer and that
is perpendicular to the clay surfaces. The rotation is performed
in such a manner that each atom with position $x_i$ is connected
by a "virtual spring" represented by a harmonic potential to its
moving reference position: $y_i = \Omega(t) (y_i^0 - y_c^0)$, where $\Omega(t)$
given below is a dimensionless matrix describing the rotation
around the axis, t is time, and $y_c^0$ is the geometrical center of the
initial reference positions (in this case, it corresponds to the
geometrical center of the small clay layer).
\begin{equation}
\ \left( \begin{array}{ccc}
cos\omega t + \nu_x^2\xi & \nu_x \nu_y \xi - \nu_z sin \omega t & \nu_x \nu_z \xi + \nu_y sin \omega t \\
\nu_x \nu_y \xi + \nu_z sin \omega t & cos\omega t + \nu_y^2\xi & \nu_x \nu_z \xi - \nu_x sin \omega t \\
\nu_x \nu_z \xi - \nu_y sin \omega t & \nu_x \nu_z \xi + \nu_x sin \omega t & cos\omega t + \nu_z^2\xi \end{array} \right)\ 
\end{equation}
where $\nu_x$, $\nu_y$, and $\nu_z$ are the components of the normalized
rotation vector, $\nu$ and $\xi = 1 - cos\omega t$.
To achieve unrestrained motion along the rotational axis and
to allow adjustment of the $d_{001}$-spacing during rotation, the
components of the potential parallel to the axis are removed.
This is done by projecting the distance vectors between the
reference and actual atomic positions onto the plane
perpendicular to the rotation axis. Thus, the final form of the
potential is
\begin{equation}
\ V^{rot} = \frac{k}{2}\sum_{i=1}^N {\Omega(t)(y_i^0 - y_c^0) - (x_i - x_c)
   - {{\Omega(t)(y_i^0 - y_i^0) - (x_i - x_c)}\nu}\nu}^2
\end{equation}
where k is a spring constant and $x_c$ is the geometrical center of
the group (the small clay layer), and $x_i$, $y_i^0$, $y_c^0$, $\nu$, $\omega$, and t are defined above. The details of the implementation can be found
elsewhere.\cite{Kutzner2011} The k value was chosen to be 100 kJ/mol.nm2,
which is low enough to ensure smooth transitions between
atomic positions. To perform the enforced rotation, the clay
systems were first equilibrated at various compositions of the
interlayer species at $\theta$ = 0\textdegree. Rotation was induced around the
axis at an angular rate of $\omega$ = 0.01\textdegree /ps over 1.2 ns with a step
size of 0.5 fs. This produces a 12\textdegree\space rotation of the clay layers
passing various Moire patterns. Additional simulations were
conducted using $\omega$ = 0.001\textdegree\space /ps to study the dependence of
potential energy on the angular rate.

\section{Results and Discussion}
Table 1 compares the $d_{001}$-spacings of dry M-MMT (M = metal
ion) computed in this work with those from DFT calculations
\begin {table}[h!]
\begin{center}
\caption {$d_{001}$-Spacing Parameters for Dry M-MMT Systems} 
\begin{tabular}{l*{3}r}
ion              & this work & DFT(vdW-TS) & exptl\\
\hline
$Na^+$           & $9.52 \pm 0.02$;$(9.43 \pm 0.01)$ & $9.47^b$ & $9.6^c ; 9.6^d$\\      
$Ca^{2+}$        & $9.56 \pm 0.02$;$(9.42 \pm 0.01)$ & $9.33^b$ & $10.0^c ; 9.6^d$\\
$K^+$            & $10.24 \pm 0.03$;$(9.83 \pm 0.02)$ & $9.88^b$ & $10.0^c ; 10.0^d; 10.2^e$\\
\end{tabular}\par
\end{center}
$^a$ The values in parentheses are computed using a $8 \times 4 \times 4$ supercell.
$^b$Ref \cite{Voora2011}. $^c$Ref\cite{Ferrage2005}. $^d$Ref\cite{Abramova2007}. $^e$Ref\cite{Morodome2011}.
\end{table}
\newline
\begin{figure}[h!]
\centering
\includegraphics[scale=0.5]{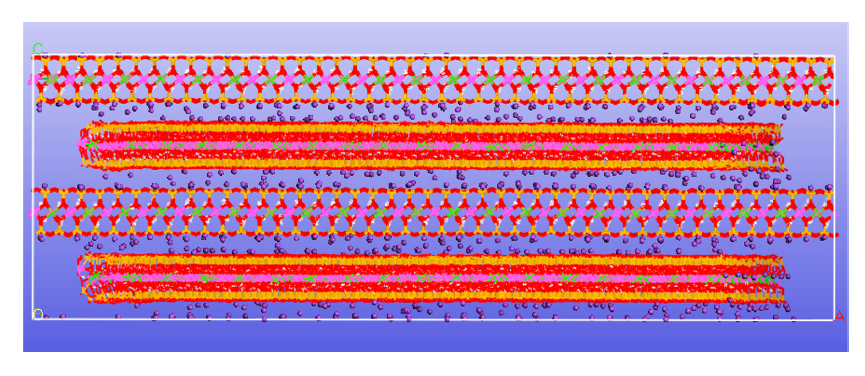}
\caption{Equilibrium structure of the dehydrated Na-MMT system at $\theta$ = 6\textdegree . Color designation: red balls, oxygen; yellow, silicon; cyan, aluminum; green, magnesium; purple, sodium ion; white, hydrogen.}
\end{figure}
with dispersion corrections\cite{Voora2011} and from experiment as inferred
from the relative proportions of the different layer types as a
function of relative humidity.\cite{Ferrage2005,Abramova2007,Morodome2011} There is an overall good
agreement between the computed and measured data, although
the experimental $d_{001}$-spacing values are prone to large
uncertainties because the experimental samples were prepared
by dehydrating humidity-exposed clays, which may have caused
the original structure not to be preserved.\cite{Boek1995,Tambach2004} Figure 3 depicts
equilibrated structures of dry Na-MMT at $\theta$ = 6\textdegree\space (as an
example). The disordering is represented as alternating clay
layers rotated around an axis perpendicular to the internal clay
surfaces. In a previous paper\cite{Myshakin2013} using the same force fields,\cite{Cygan2004,Cygan2012} we demonstrated that the expansion of the $d_{001}$-spacing of Na-MMT upon intercalation of water closely reproduces the experimental dependence of the $d_{001}$-spacing as a function of
interlayer water content.\cite{Fu1990} The $d_{001}$-spacing displays plateaux
corresponding to a stable hydration states.\cite{Fu1990} The predicted
$d_{001}$-spacings are also consistent with recently reported data on
Na-montmorillonite and Na-hectorite.\cite{Krishnan2013,Cygan2012,Morrow2013}\newline
Figures 4 and 5 display, respectively, the relative changes in the potential energy and the $d_{001}$-spacing as a function of X-Y interlayer compositions and the $\theta$ angle. These data were obtained using atomic positions constrained to predetermined $\theta$ angles. The upper part of each figure shows data without carbon dioxide, and the lower part reports results with intercalated $CO_2$. The interlayer spacing changes within a narrow 0.2 \AA\space 
range (except the Ca-MMT system at the 5.2
composition, which displays a variation of up to 0.35 \AA) upon
rotation. The differences between the calculated enthalpy and
potential energy changes are negligible, so in the subsequent
analysis, we use the potential energy as the quantity to analyze
energy dependencies.
\\In general, the largest changes in the potential energy and
structural parameters occur during rotation from 0\textdegree\space to 3\textdegree, and a further increase in $\theta$ perturbs the systems to a lesser extent(except for the 0.2 composition). This might be related to the fact that the ratio of the unit cell area to the concentric circle
area does not depend on $\theta$ (see Computational Details).
Consequently, the numbers of distorted and undistorted
cavities remain fixed. Introduction of $Ca^{2+}$ and $K^+$ counterions
produce more noticeable changes in the energy and interlayer
distance than found for the $Na^+$ ion. Thus, cations having larger
ionic radii and charge affect the potential energy and the
interlayer distance more. Below, we consider the dehydrated,
hydrated, and clay systems with intercalated carbon dioxide
separately, paying attention to the energetic and structural
changes and the density distributions of the interlayer species as
a function of $\theta$.
\newline
\begin{figure}[h!]
\centering
\includegraphics[scale=0.5]{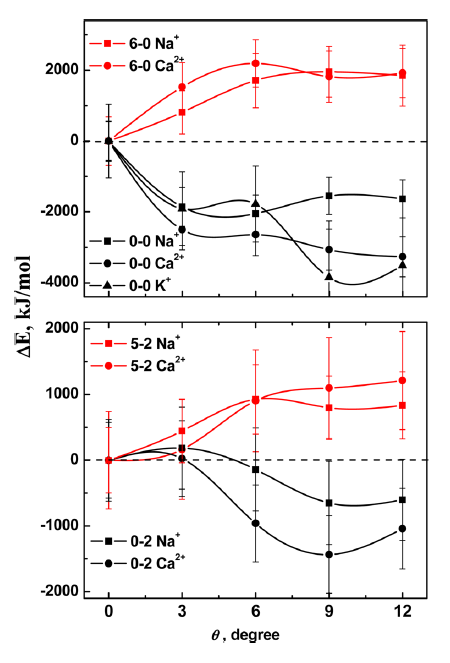}
\caption{Relative potential energy change in M-MMT systems as a
function of interlayer compositions and $\theta$.}
\end{figure}
\newline
\begin{figure}[h!]
\centering
\includegraphics[scale=0.5]{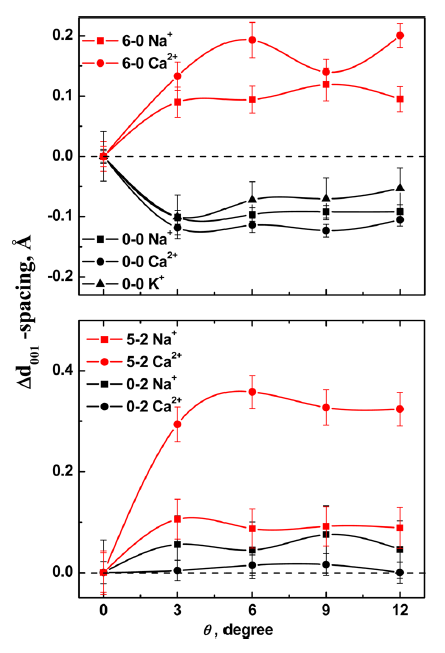}
\caption{Relative $d_{001}$-spacing change in M-MMT systems as a
function of interlayer compositions and $\theta$.}
\end{figure}

\subsection{Rotation of Dry Clay Systems}
For the dehydrated clay systems, the potential energies for the twisted structures are lower than for the structure at $\theta$ = 0 \textdegree\space (Figure 4). The
corresponding $d_{001}$-spacings also demonstrate decreased values
compared with the $\theta$ = 0 \textdegree\space case (Figure 5). The figures depict
the intriguing result that the energy and the $d_{001}$-spacing are
lower for the rotated dehydrated Na- and Ca-MMT systems
than for the clay structures at $\theta$ = 0\textdegree\space . The ions can be
coordinated equidistantly to negatively charged basal oxygens
of ditrigonal rings in perfectly juxtaposed clay sheets (Figure
2a). Such coordination provides optimal interaction, resulting
in lowering of total energy. Because rotation causes reduction
in a number of those cavities, as evidenced by the formation of
the Moire patterns (Figure 1), the total energy of the system
would have been expected to increase. Before addressing this
issue, it is important to mention that the dry clay systems
considered in this study are idealized structures. In reality, a
smectite sample in the 0W hydration state is expected to have
residual water bound to interlayer ions.\cite{Ferrage2005} To check the effect of
trace amounts of water on the potential energy and interlayer
distance, a set of simulations using 0.25 water molecule per unit
cell of Na-MMT was conducted. These simulations gave
energies and $d_{001}$-spacings of rotated structures at $\theta > 0$\textdegree\space lower
than the corresponding values at $\theta$ = 0\textdegree. Thus, even with a
small amount of water present, the distribution of interlayer
ions still controls the trends in relative energy and $d_{001}$-spacing
depicted in Figures 4 and 5. We now turn to the results for the
fully dehydrated Na- and Ca-MMT systems.
\newline
\begin{figure}[h!]
\centering
\includegraphics[scale=0.6]{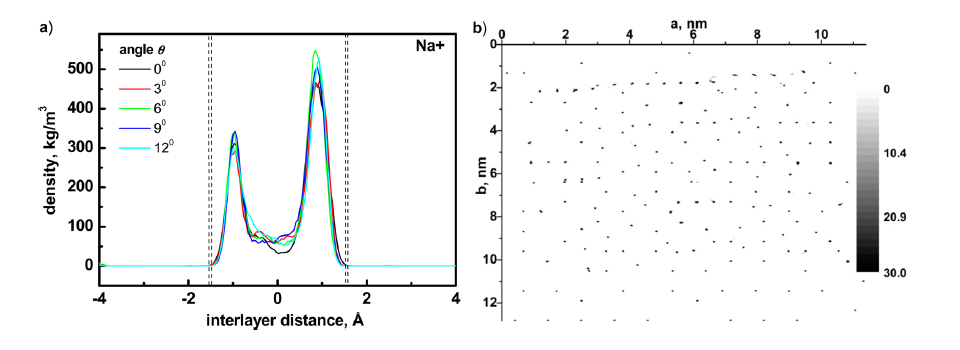}
\caption{(a) Density profiles showing sodium ion distribution along the distance perpendicular to the internal surfaces of Na-MMT with the 0-0
composition at various θ values. Double dashed lines designate the range of positions of basal planes as a result of rotation. Profiles obtained by averaging over 5 ns of simulation time. (b) Density map (density in number/$nm^3$) showing the sodium ion distribution in the interlayer projected on a plane parallel to the internal surfaces of Na-MMT with the 0-0 composition at $\theta$ = 6\textdegree. Results obtained by averaging over 5 ns of simulation time.}
\end{figure}
\\The atomic density profiles for the 0.0 composition
depicted in Figure 6a indicate that $Na^+$ ions (and $Ca^{2+}$, not shown) display two asymmetric peaks located near the basal
planes designated by the dashed lines. The asymmetry is caused
by different sizes of the alternating clay layers, as described in Computational Details. This trend does not depend on the
initial positions of the ions. The analysis of the distances
between the ions and the basal oxygens shows that ions
associated with the density peaks are preferentially coordinated
at the middle of the ditrigonal rings within 2.9 \AA\space (estimated for sodium ions) from three and more basal oxygen atoms, thus
meeting the criterion used to assign an ion as adsorbed.\cite{Myshakin2013} This coordination does not depend on initial positions and is
reproduced with independent simulations. Morrow et al.\cite{Morrow2013} also found that in dehydrated samples of Na-hectorite, the $Na^+$ ions lie at the center of the hexagonal rings on the sheet to which they are closest. Similar observation was made in the MD
simulations of Na-rectorate.\cite{Jinhong2012} The number of ions is 0.75 (two times smaller in the case of Ca-MMT) per four ditrigonal rings (two on each surface) in a unit cell so that there is no
competition for a coordination place that would otherwise force
them to stay close to the interlayer center plane. At the middle
of the ditrigonal rings, the counterbalancing ions are strongly
electrostatically bound to surrounding basal oxygens. The
equilibration step allows the interlayer ions to find their
energetically favorable configurations at each $\theta$ value
considered. Such a process might mimic the situation occurring
during dehydration of clay samples upon heating. Thus, under
(geo)mechanical stress, the interlayer ions in smectite clay
minerals slowly losing water would have sufficient time to
adjust their positions and promote rotational disordering at
least within the range of $\theta$ values considered in this work.
\\Figure 6b depicts the two-dimensional density map for the
sodium ions in the interlayer. The map was obtained by
scanning the interlayer space to obtain density distributions in
planes parallel to the clay surfaces with a step size of 0.1 \AA.
Then the distributions were projected on a plane and averaged
for 5 ns of simulation time. The map shows that the sodium
ions are localized with respect to their XY coordinates (similar
distributions were obtained for the other $\theta$ angles). This,
together with the density profile data, also confirms ion
localization near the surfaces.
\newline
\begin{figure}[h!]
\centering
\includegraphics[scale=0.5]{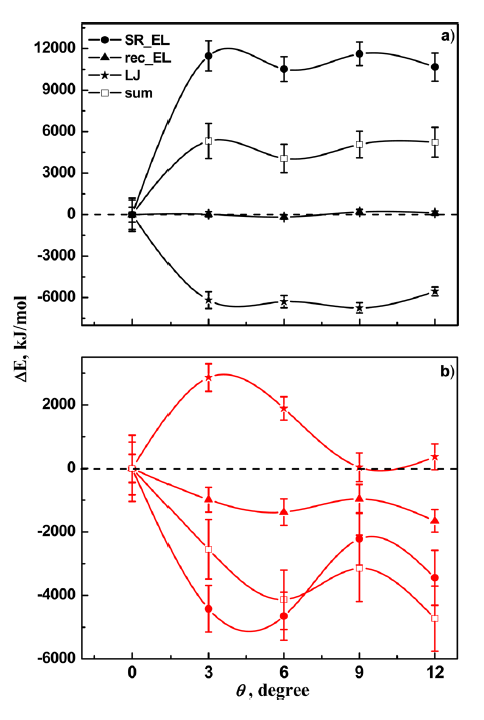}
\caption{Comparison of real-space electrostatic (SR\_EL), reciprocal
electrostatic (rec\_EL), Lennard-Jones (LJ) contributions and their
total sum (sum) as a function of $\theta$ for the (a) 6-0 and (b) 0-0
compositions of Ca-MMT. }
\end{figure}
\newline
\begin{figure}[h!]
\centering
\includegraphics[scale=0.5]{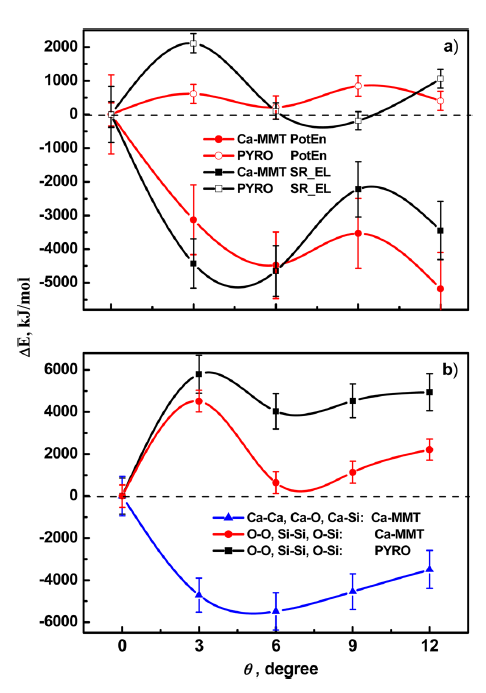}
\caption{Comparison of (a) relative potential energies (PotEn) and
real-space electrostatic contributions (SR\_EL) and (b) atomic pairwise
electrostatic contributions to relative potential energies in dry Ca-
MMT and pyrophyllite with the 0-0 composition at various $\theta$ values.}
\end{figure}
\\To gain insight into the interactions responsible for the
computed energy trend, contributions from the various terms
in the force field were extracted and analyzed. Figure 7 reports
short-range (real-space sum accounting for pairwise interactions
within a sphere of 11 \AA, the cutoff radius), reciprocal
space electrostatics and LJ contributions into the relative
potential energies as a function of $\theta$ for the 6-0 (upper part)
and 0-0 (lower part) compositions. The "position-restraining"
and "long-range dispersion correction" (a minor modification
to the Lennard-Jones terms to remove the noise caused by
cutoff effects) terms provide insignificant contributions to the
relative potential energies at different $\theta$ values (it should be mentioned that the weighted histogram analysis method\cite{Gromacs} would be valuable for removing the bias due to the restraint and for estimating free energies as a function of $\theta$, but it requires a special effort to be applied for these clay systems and was not considered in this work). For the hydrated system, rotation of the small layer relative to the large one leads to a large electrostatic destabilization, whereas the LJ contribution is energetically favorable. On the other hand, for the dehydrated system, the electrostatics favors the rotated structures with the LJ contribution being positive. Comparing the relative influence
of the different terms in the force field, it is seen that short range electrostatic contributions are the most important for the
overall potential energy change as a function of $\theta$ for both hydrated and dry MMT systems.
\newline
\begin{figure}[h!]
\centering
\includegraphics[scale=0.5]{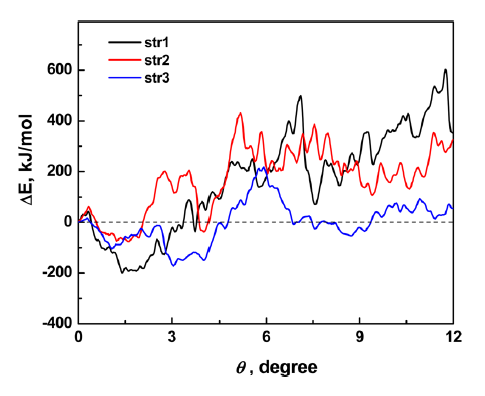}
\caption{Potential energy changes during enforced rotation for three
independently equilibrated Na-MMT systems (strs 1-3) with the 0-0
composition.}
\end{figure}
\\To estimate the role of electrostatic contributions involving
interlayer ions, simulations were also carried out for
pyrophyllite, which does not have octahedral and tetrahedral
substitutions in TOT clay layers and, thus, bears neutral layers
without interlayer ions. Figure 8 (upper part) depicts the
relative potential energies and the relative real-space electrostatic contributions for pyrophyllite and Ca-MMT. The key
atomic pairwise electrostatic contributions involving the ions
and atoms of basal surfaces are depicted in the lower part of
Figure 8. In contrast to Ca-MMT, the rotation of dry
pyrophyllite is an energetically demanding process, and for
both minerals, the electrostatic contribution to the relative
potential energies determines the overall decrease/increase in
the potential energies with rotation. The atomic pairwise
contributions from the basal oxygen and silicon atoms provide a
similar destabilizing impact for the rotated pyrophyllite and Ca-
MMT systems. However, in the case of Ca-MMT, the
electrostatic contribution involving interlayer ions induces a
large stabilization upon rotation. Therefore, the overall decrease of the potential energy of nonhydrated MMT systems with
rotation is primarily caused by interactions of interlayer ions
with atoms of the clay surfaces.
\\We now turn our attention to exploring the energetic and
structural changes during rotational motion of the small clay
layer at a constant angular velocity using the enforced rotation
approach. Figure 9 depicts the evolution of the potential energy
as a function of $\theta$. Three independently equilibrated Na-MMT systems with the 0-0 composition at $\theta$ = 0\textdegree\space were used as starting points to initiate simulations, allowing for rotational movement of the small layer relative to the large one. The initial motion from $\theta$ = 0\textdegree\space to $\theta$ = 0.1-0.2\textdegree\space requires overcoming a potential barrier of around 25.50 kJ/mol. The barrier is associated with a deviation of the system from its equilibrium while the Moire pattern is not yet developed (for the size of the clay system employed). After that, as soon as the Moire pattern begins to form, further rotation leads to a decrease in the energy (and in the $d_{001}$-spacing, not shown) up to $\theta$ = 1.5-2\textdegree.
The energy (and $d_{001}$-spacing) then increases upon further
increase in $\theta$ (Figure 9). Analysis of the distribution of the interlayer ions during the enforced rotation shows that the
rotational motion does not induce rearrangement of the ions
adsorbed at the centers of ditrigonal rings. The energy
decreases for $\theta$ = 0.2 - 2 \textdegree\space because the ions remain close to equilibrated positions at $\theta$ = 0\textdegree . However, rotation to larger
angles ($>3$\textdegree) causes more unfavorable ion-ion interactions between ions adsorbed at the opposite internal clay surfaces. Those interactions counterbalance the energy decrease that would otherwise result because of rotation. As a result, further disordering beyond 2-3\textdegree\space would be significantly retarded.
\newline
\begin{figure}[h!]
\centering
\includegraphics[scale=0.6]{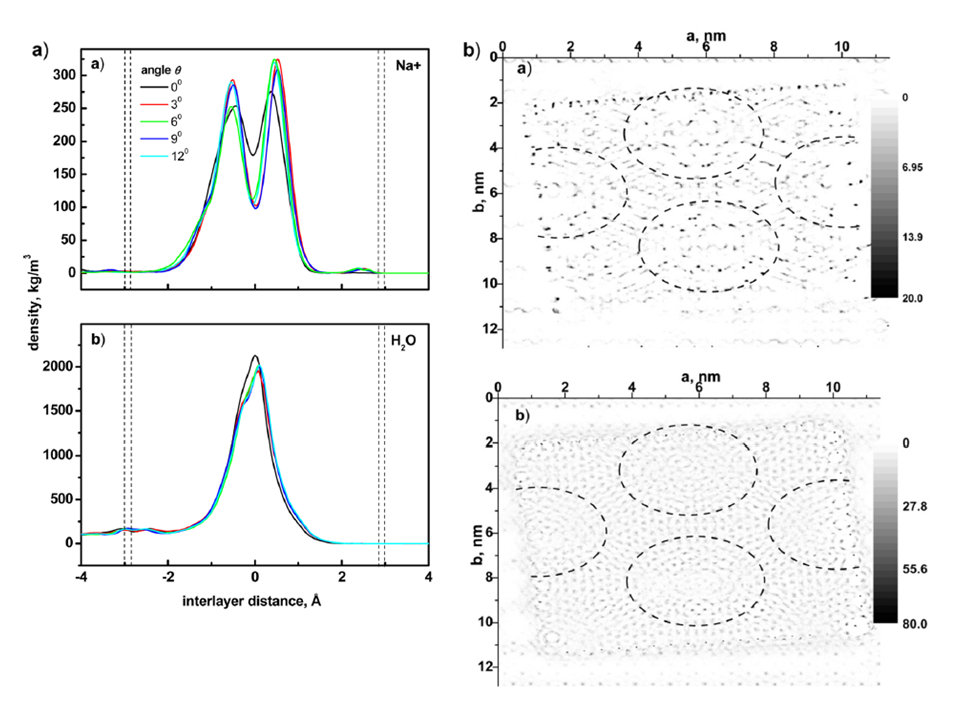}
\caption{(a; left) Density profiles showing (a) sodium ion and (b) water molecule distributions along the distance perpendicular to the internal surfaces of Na-MMT with the 6-0 composition at various $\theta$. Double dashed lines designate the range of positions of basal planes as a result of rotation. Profiles obtained by averaging over 5 ns of simulation time. (b; right) Density maps (density in number/$nm^3$) showing (a) sodium ion and (b) water molecule distributions in the interlayer projected on a plane parallel to the internal surfaces of Na-MMT with the 6-0 composition at $\theta$ = 6\textdegree. Results obtained by averaging over 5 ns of simulation time. The Moire pattern is highlighted with dotted circles.}
\end{figure}
\\The simulated rotational motion models the disordering
occurring in dehydrated or nearly dehydrated smectite
minerals. The energy trends obtained using the enforced
rotation are different from those found with the position
constraints. The differences originate from the fact that the
latter approach allows relaxation of the ion positions at fixed
rotational angles, whereas in the former, the ions remain
adsorbed at positions corresponding to $\theta$ = 0\textdegree. The rate of the motion, which is limited by computational cost, is apparently much greater than that occurring in nature and proceeds under nonequilibrium conditions. To explore rotational motion at a reduced angular speed, simulations were performed using a rotational velocity of 0.001\textdegree /ps. The predicted trends in energy and $d_{001}$-spacing remain similar to those discussed above. This implies that under mechanical stress, nearly dry smectite clay minerals would be susceptible to rotational disordering within the narrow $\theta$ range.
\\To explore an effect of the Moire patterns on the potential
energy of dry smectites, simulations with backward enforced
rotation of MMT systems were carried out for clay structures at
nonzero $\theta$ values. Figure S1 (Supporting Information) collects potential energy curves computed during backward rotation for Na-MMT structures with interlayer ions equilibrated using the position restraining method at $\theta$ = 3\textdegree, 6\textdegree, and 9\textdegree. The curves develop distinct peaks when rotation of the small layers passes the position of perfectly juxtaposed clay layers at $\theta$ = 0\textdegree. The
deviation from that position in both clockwise and counterclockwise directions leads to a decrease in energy. Thus,
formation of the rotational patterns of adjacent clay surfaces
induces a decrease in the potential energy (and $d_{001}$-spacing) of the dehydrated smectites.
\newline
\begin{figure}[h!]
\centering
\includegraphics[scale=0.6]{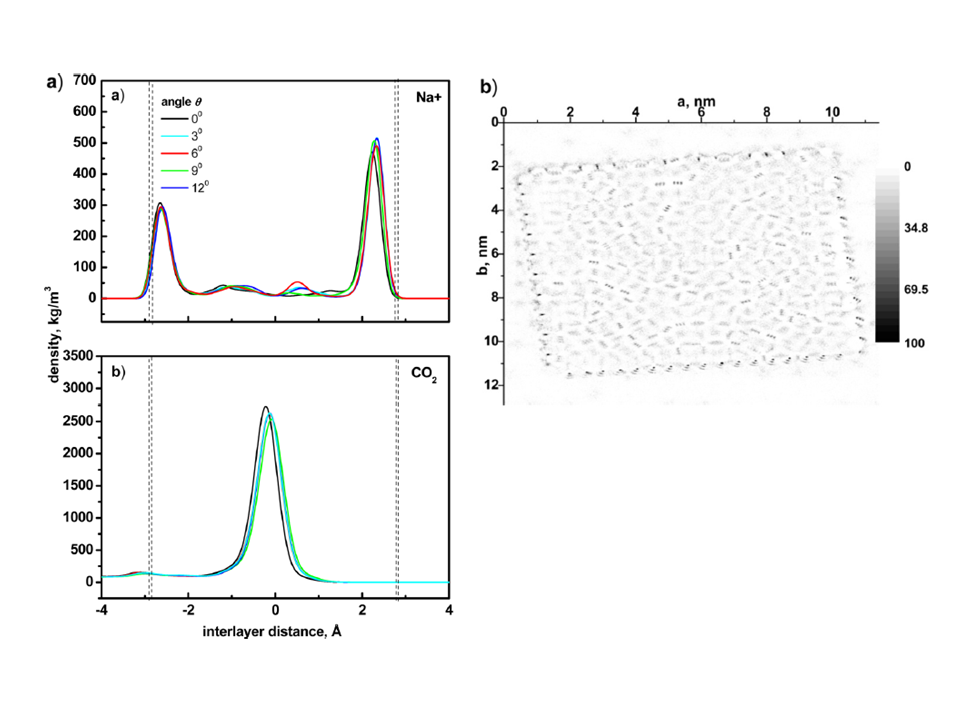}
\caption{(a; left) Density profiles showing (a) sodium ion and (b) carbon dioxide molecule distributions along the distance perpendicular to the internal surfaces of Na-MMT with the 0-2 composition at various $\theta$ values. Double dashed lines designate the range of positions of basal planes as a result of rotation. Profiles obtained by averaging over 5 ns of simulation time. (b; right) Density maps (density in number/$nm^3$) showing carbon dioxide molecule distributions in the interlayer projected on a plane parallel to the internal surfaces of Na-MMT with the 0-2 composition at $\theta$ = 6\textdegree. Results obtained by averaging over 5 ns of simulation time.}
\end{figure}
\subsection{Rotation of Hydrated Clay Systems}
According to Figures 4 and 5, rotation of the clay layers of hydrated Na- and Ca-MMT requires energy and is accompanied by expansion of the interlayer distance. The overall trends in energies, $d_{001}$-spacings, atomic density distributions, and radial distribution functions are similar for the 2-0, 4-0, 6-0, and 8-0 compositions. We now examine in more detail the results of the simulations for the 6-0 composition, which provides a water monolayer and a $d_{001}$-spacing equal to $12.10 \pm 0.03$ \AA \space($\theta$ = 0\textdegree). That equilibrated spacing falls within the experimentally determined range of the 1W hydration state (11.5-12.5 \AA)\cite{Fu1990} and corresponds to the calculated stable hydration state.\cite{Marry2002,Cygan2004,Whitley2004,Smith2004} As a function of $\theta$, Figure 7 (upper part) displays the various contributions to the relative potential energy. The figure shows that electrostatics plays the primary role in the overall increase of the relative potential energy upon rotation. Detailed analysis of the pairwise contributions relative to the zero-degree system shows that the electrostatic contributions from the basal oxygen-basal silicon, ion-basal oxygen, and water-water interactions contribute to the overall increase in the relative potential energy. Hence, the electrostatic contribution due to the water network is less stabilizing for rotated than nonrotated clay layers (Supporting Information Figure S2).
\\As a function of $\theta$, Figure 10a displays the density profiles of the sodium ions and water molecules (the profile does not distinguish individual atoms in the molecules) in the interlayer for the 6-0 composition. The density profile of water develops a maximum at the interlayer center with a low shoulder caused by coordination of water molecules to the clay layer edges. The density profile of the sodium ions is characterized by two peaks that occur close to the interlayer center. Interestingly, beginning with the $\theta$ = 3\textdegree\space system, the profiles develop more
pronounced peaks relative to $\theta$ = 0\textdegree. This indicates that upon rotation, ions are driven away from the interlayer center toward the basal planes. The process might be related to the less stabilizing impact of water-water electrostatics to the potential energy of the rotated systems. 
\\Figure 10b shows the 2D density maps of the ions and water molecules for the 6-0 composition. Examination of Figures 1, 10b, and Supporting Information S3 (given for $\theta$ = 0\textdegree) shows that the $H_2O$ molecules and sodium ions create patterns reminiscent of the Moire patterns for nonzero $\theta$ values. This striking feature indicates that rotational disordering affects the structural arrangement of the interlayer species for hydrated
montmorillonite. Similar results are obtained from analysis of the simulation data for Ca-MMT. The density distribution map also demonstrates that the ions actively explore the interlayer space in the XY planes parallel to the clay surfaces. This is distinctly different from the distribution of the ions at the 0-0 composition that revealed sharp localization in the XY plane (Figure 6b). This is a consequence of the fact that solvation by water molecules provides greater ion mobility in the interior.
\\In the simulations of Na-MMT with the 6-0 composition
using enforced rotation, the energy and $d_{001}$ spacing are higher for rotated structures than for the system with $\theta$ = 0\textdegree\space (not shown). The energy curve reaches a plateau and stays relatively constant during the rotation starting from $\theta > 2-3$\textdegree.
\\So far, we have considered the hydrated clay systems with
water monolayer exposed to the Moire. patterns of clay sheets. Supporting Information Figure S4 shows the potential energy changes as a function of $\theta$ computed using both the position constraining and enforced rotation methods for the Na-MMT with the 14-0 composition. This composition produces a bilayer configuration of water molecules (as confirmed by the density profile) in the interlayer with the equilibrated $d_{001}-spacing$ equal to 14.67 $\pm$ 0.03 \AA, the value that falls into the experimentally determined 2W hydration state range (14.5-15.5 \AA)\cite{Fu1990} and close to the calculated stable hydration state.\cite{Marry2002,Cygan2004,Whitley2004} In this case, both methods predict that rotation from $\theta$ = 0\textdegree\space is an endothermic process accompanied by the
increase in the $d_{001}$-spacing. The density profiles (not shown) support the preferential location of the sodium ions near the center of the interlayer space and their solvation by water molecules. This is in line with the recent DFT-based molecular dynamics and Monte Carlo simulations of optimal ion positions in the interlayer space of hydrated montmorillonite.\cite{Suter2012} In contrast to the system with the 6-0 composition, the calculated density maps indicate that ion distribution is relatively unaffected by the Moire pattern; the same is true for the water molecules, although they develop a cage-like pattern owing to the presence of an adjacent basal surface (Supporting Information Figures S5.S7). The pictures provide the distribution maps computed for the ions and separately for water layers in a bilayer configuration at $\theta$ = 6\textdegree\space as an example
(the distributions for the other angles are similar). Thus, the Moire pattern exerts little influence on the distribution of the interlayer species for clay expanded into the 2W (and presumably also higher) hydration state.\newline
\begin{figure}[h!]
\centering
\includegraphics[scale=0.8]{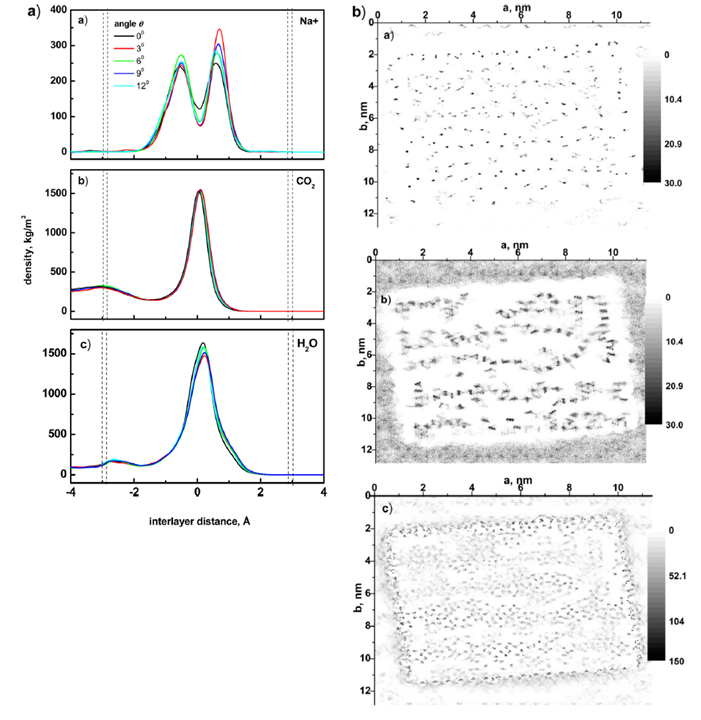}
\caption{(a; left) Density profiles showing (a) sodium ion, (b) carbon dioxide, and (c) water molecule distributions along the distance perpendicular to the internal surfaces of Na-MMT for the 5-2 composition at various $\theta$. Double dashed lines designate the range of positions of basal planes as a result of rotation. Profiles obtained by averaging over 20 ns of simulation time. (b; right) Density maps (density in number/$nm^3$) showing (a) sodium ion, (b) carbon dioxide, and (c) water molecule distributions in the interlayer projected on a plane parallel to the internal surfaces of Na-MMT with the 5-2 composition at $\theta$ = 6\textdegree. Results obtained by averaging over 20 ns of simulation time.}
\end{figure}

\subsection{Rotation of Clay Systems with Intercalated $CO_2$}
As seen from Figures 4 and 5 the potential energy of the clay systems with 0-2 composition undergoes a slight increase as $\theta$ increases from 0\textdegree\space to 3\textdegree, then decreases for higher rotational angles relative to the 0\textdegree\space case. The corresponding $d_{001}$-spacing experiences a marginal expansion upon rotation. Under experimental conditions, intercalation of carbon dioxide requires a residual amount of water present in the interlayer
space.\cite{Schaef2012,Loring2012,Giesting2012} The 0-2 composition considered here is an idealized system not likely observable experimentally. In contrast to this, rotation of the clay system with the 5-2 composition is accompanied by an increase in both the potential energy and the $d_{001}$-spacing (Figures 4 and 5).
\\Figure 11a and b depicts the density profiles for $Na^+$ ions and carbon dioxide molecules and density distribution maps for carbon dioxide molecules in Na-MMT with the 0-2 composition. The resulting profiles of the sodium ions are reminiscent of those found for the 0-0 composition (Figure
6a). The density map of the $Na^+$ ions (not shown) is also similar to that depicted in Figure 6b. Krishnan et al.\cite{Krishnan2013} reported a detailed analysis of the dynamics of sodium ions and carbon dioxide molecules in the interlayer space of Na-MMT. They found that the ions primarily adsorbed at the surfaces, with
some ions being located in the interior. Similar results are obtained in this work and depicted in Figure 11a, showing two minor peaks near the interlayer center. The density profile of
carbon dioxide displays a maximum at the middle of the interlayer (Figure 11a). Thus, carbon dioxide and the sodium ions avoid each other, consistent with previous data.\cite{Myshakin2013,Krishnan2013}
\\Comparison of the density maps for the nonzero $\theta$ value (Figure 11b) and for $\theta$ = 0\textdegree\space (Supporting Information Figure
S8) has revealed that the $CO_2$ distribution is affected by the Moire patterns formed by the basal surfaces. However, the distribution is less structured, to be recognized as the Moire pattern like that depicted in Figure 10b for water molecules.
Similar trends were found for the Ca-MMT system.
\\Figure 12 reports the density profiles at various $\theta$ values and density distribution maps of the interlayer species for the 5-2 composition of Na-MMT. The sodium ion profile displays two
peaks located in the interior of the interlayer space similar to the profile for the 6-0 composition (Figure 10a), suggesting that the ions become preferentially solvated by water molecules.
Moreover, in line with Figure 10a, rotation of the clay sheets causes a slight increase in the separation of the peaks in the $Na^+$
distribution. For the 5-2 composition, the $CO_2/H_2O$ mixture forms a monolayer correlating with the maxima of the density profiles. At the positions of the layer edges, the carbon dioxide profile develops a distinct broad shoulder and the density profile of water demonstrates a small peak (Figure 12a). The density distribution map of carbon dioxide shows elevated density in the interstitial space and at the edges of the small
layer that gives rise to the shoulder seen in the density profile.
\\Similarly, the density distribution map of water displays an increase in density around the layer edges, although to a much lesser extent (Figure 12b). Apparently, the 5-2 composition
(and 5-1, not shown) creates a mixture with a supersaturated concentration of carbon dioxide, and during equilibration, the excess of $CO_2$ moves from the interlayer to the interstitial space. In our earlier work, the 5-2 composition resulted in the
equilibrium $d_{001}$-spacing equal to 14.49 $\pm$ 0.02 \AA, which falls within the range of the 2W hydration state.\cite{Myshakin2013} The Na-MMT
model employed in that work provides exposure of the
interlayer species only to the internal clay surfaces (no edge effects). In this study, because the interlayer species have access to the interstitial space and edges, the equilibrium $d_{001}$-spacing
decreased to 12.17 $\pm$ 0.02 A ($\theta$ = 0\textdegree), close to the value for the
6-0 composition (see section 3-2). Interestingly, the carbon dioxide remaining in the interlayer region tends to agglomerate and form elongated clusters (conglomerates) separated by
water molecules with a residual amount of solvated $CO_2$ (Figure 12b). To explore the stability of such clusters, production simulations of Na- and Ca-MMT at $\theta$ = 6\textdegree\space were conducted up to 200 ns to monitor potential energy, $d_{001}$-
spacing changes, and density distribution maps. At the end of the simulations, the density distributions remain unchanged, as reported in Figure 12. Density maps of water molecules display
nonuniform distributions reflecting the presence of agglomerated carbon dioxide molecules in the interlayer. There are no noticeable patterns formed by water that might be connected to
the Moire patterns of basal surfaces, as found for the 6-0 composition (Figure 12b).
The hydrogen bonding between the interlayer species and the basal oxygens affects the energetic properties of clay systems. In this regard, it is instructive to explore hydrogen bond lifetimes as a function of interlayer composition and $\theta$. To
accomplish this, the approach described in ref \cite{Spoel2006} was engaged.
\\The analysis of the calculated hydrogen-bond lifetimes for water-clay, water-water, and water-carbon dioxide pairs in Na- and Ca-MMT at different $\theta$ values shows that the rotational disordering does not significantly impact the H-bond
lifetimes. The water-water and water-basal oxygen hydrogen lifetimes computed for the 6-0 and 5-2 compositions have comparable values (the water-carbon dioxide values are an order of magnitude smaller, consistent with our previous results).\cite{Myshakin2013} This means that in the interlayer, the water molecules are strongly engaged in interactions with the basal surfaces. On the other hand, in the open environment, the H-bonding
of water to a single smectite surface is weaker than H-bonds between water molecules.\cite{Zhang2012, Rotenberg2011, Marry2008}
The simulations of Na-MMT systems for the 0-2 and 5-2 compositions using the enforced rotation approach demonstrate a rise in the relative potential energy (and $d_{001}$-spacing)
upon rotation (not shown). The energy increases even for the 0-2 composition that is found to demonstrate the opposite trend using the position restraint approach. This is attributed to nonequilibrium configurations of the interlayer species during
enforced rotation. Specifically, deviation of the clay layers from $\theta$ = 0\textdegree\space induces an increase in the potential energy of the system.
To provide smoother rotation motion, the angular speed was decreased by an order of magnitude to 0.001\textdegree /ps. However, it did not reverse the trend, meaning that the 0-2 system is more
sensitive to equilibration of the interlayer species than the 0-0 system, which displays a decrease in the energy and the interlayer distance at the small $\theta$ values (0-2\textdegree) (Figure 9).
\\In the discussion above, it was assumed that interlayer species possess the ability to leave the interlayer and enter the interstitial space between clay particles. This is possible if clay generally exists as particles of a limited size and thickness. In the series of works by Nadeau et al.\cite{Nadeau1984,Nadeau1985} it was found using XRD and TEM measurements that naturally occurring interstratified illite-smectite and chlorite-smectite and pure smectite are
mixtures of thin particles of finite sizes rather than a continuous crystal phase. Interestingly, the samples of illite-smectite and chlorite-smectite display rotational turbostratic electron
diffraction patterns, presumably as a consequence of clay particle agglomerates.\cite{Nadeau1984,Nadeau1985} For Wyoming montmorillonite, the samples show a predominance of clay particles a few thousand angstroms in size but only 10-20 \AA\space thick. A mobile phase migrating through a geological formation enriched with
swelling clays can fill the interstitial space between the clay particles and, thus, be exposed to external mineral interfaces.
\\Carbon dioxide molecules can become trapped in the
interlayer in amounts exceeding an expected $CO_2$ solubility at the prevailing P-T conditions. It has been shown experimentally that exposure of dry sc$CO_2$ to montmorillonite in the $>$ 2W hydration state may result in a collapse of the d-spacing to
that of the 1W state.\cite{Ilton2012,Schaef2012} During that process, a portion of the water molecules leaving the interlayer can be substituted with
carbon dioxide molecules. A significant portion of (wet) sc$CO_2$ might remain in the interstitial region between clay particles.
Recently, MD simulations were used to study the exchange of water and counterbalancing ions between the micropores and clay interlayers in Na-montmorillonite with open [010]
edges.\cite{Rotenberg2007} It was found that for water content at the 2W hydration state, the exchange proceeds practically without a barrier for water and the ions. Energy barriers for exchange of
carbon dioxide, water, and ions between the interlayer and (wet) sc$CO_2$ in the interstitial space and pores are unknown. A study aiming at an estimation of those barriers would be a
valuable addition to our understanding of the mechanism of carbon dioxide interaction with swelling clay minerals. 
\\Another scenario involves intercalation of $CO_2$ during expansion of swelling clay minerals that might be at fractional hydration states\cite{Morrow2013,Tenorio2010,Tambach2006} (although it is generally believed that smectite samples exist as a set of "quantized" hydration states, i.e., (0W, 1W, 2W, 3W))\cite{Ferrage2005,Sato1992,Ferrage2007} Experimentally, it was found
that a residual amount of water is required for successful intercalation of $CO_2$ in the interlayer space that expands until the $d_{001}$-spacing corresponds to that of the 1W hydration
state.\cite{Giesting2012,Giesting2012a} Interaction of Na- and Ca-exchanged-MMT samples at the 2W hydration state with variably wet sc$CO_2$ can lead to swelling to the $d_{001}$-spacing equal to the 3W state.\cite{Ilton2012,Schaef2012} As
mentioned in the Introduction, the largest expansion occurs for MMT samples at a sub-1W hydration state and is accompanied by an increase in the $d_{001}$-spacing from 11.3 to 12.3 \AA\space after exposure to gaseous $CO_2$\cite{Giesting2012} or to 12.1 \AA\space after interaction with anhydrous sc$CO_2$.\cite{Ilton2012,Schaef2012}
\\The formation of $CO_2$ conglomerates trapped in the
interlayer is supported by Schaef et al.\cite{Schaef2012} and Rother et al.,\cite{Rother2012} who indicated that $CO_2$ does not displace $H_2O$ when entering the sub-1W interlayer but "rather makes room by pushing the structural units apart", and by simulation data of Yong and
Smith,\cite{Young2000} who reported the interlayer of Sr-MMT showing water molecules clustering around the ions and away from substitutions in the octahedral layer and formation of unoccupied regions. That unoccupied space might be filled in by another species such as carbon dioxide in a two-step
mechanism ("prop and fill") implying the existence of (meta)stable partially filled hydration states. In our work, we found conglomerate formations for both Na- and Ca-MMT clay systems, suggesting that $CO_2$ can be a filler to stabilize swelling clays during expansion. It is also important to realize
that hydration energies of interlayer ions, isomorphic substitutions in octahedral, and especially tetrahedral layers would be leading factors determining trapping of $CO_2$ as
conglomerates in smectites minerals. In the models employed in this work, only substitutions in the octahedral layer were made, so the interlayer species exposed to the basal surface were indirectly affected by charge imbalance brought by the magnesium for aluminum substitutions.

\section{Conclusions}
This study has shown that rotational disordering, a common naturally occurring phenomenon, affects the distribution of ions, water, and carbon dioxide molecules in the interlayer of swelling clays. The computed density maps reveal that the interlayer species in a monolayer configuration follow the
rotational Moire patterns formed by the basal surfaces of adjacent clay layers. The simulations indicate that rotational disordering of hydrated montmorillonite and montmorillonite
with intercalated water and carbon dioxide is an energetically demanding process, as found using the position constraining and enforced rotation approaches for $\theta$ = 0-12\textdegree. For all
compositions considered, the potential energy demonstrates a tendency to reach a plateau for $\theta$ = 6-12\textdegree\space that is attributed to a
fixed number of undistorted/distorted cavities for Moire patterns. Turbostratically stacked clay layers with intercalated water and water/$CO_2$ also experienced expansion of interlayer space by 0.1-0.2 \AA, depending on the nature of interlayer ions.
\\Turbostratic dry and nearly dry montmorillonite systems are predicted to be more stable than the nonrotated system. Rotation is accompanied by a decrease in the $d_{001}$-spacing by $\sim$0.1 \AA. This process is explained in terms of favorable interactions of interlayer ions adsorbed at the clay surfaces. During equilibration, the ions find optimal positions in the interlayer, causing a decrease in the potential energy and the $d_{001}$-spacing. In a geological formation, this process may be anticipated during slow dehydration of expandable clays under geomechanical stress. Under such conditions, the clay layers would be prone to rotational disorder and to become turbostatically stacked in the presence of external forces
shifting the clay particles. The calculation using enforced rotation of clay layers shows that perfectly oriented dehydrated montmorillonite has only a limited range ($\theta$ = 0-2\textdegree) for
rotational disordering, and further rotation would be energetically restricted.
\\The results of the simulations have also shown that $CO_2$ might be trapped in the interlayer of hydrated montmorillonite in an amount exceeding its solubility in water at prevailing P-T conditions in subsurface geological formations. This is possible
because carbon dioxide conglomerates become trapped in the interlayer and are surrounded by water molecules with solvated $CO_2$. Thus, the expandable clay layers provide a confining environment for such carbon dioxide retention. This trapping mechanism could be important in estimations of a storage
capacity for selected geological sites.

\section{Supporting Information}
Eight figures describing potential energy change, electrostatic contributions of various atom pairs, 2D density maps, relative potential energy and $d_{001}$ spacing change in Na-MMT systems. This material is available free of charge via the Internet at
http://pubs.acs.org.

\bibliography{ref.bib}

\end{document}